\documentclass{mn2e}
\input{epsf}

\def \vt{\vartheta}

\def \farcm{\hbox{$.\!\!^{\prime}$}}

\title[Mass profiles from weak lensing]{How well can we determine
cluster mass profiles from weak lensing?}

\author[Henk Hoekstra]{Henk Hoekstra\\
Canadian Institute for Theoretical Astrophysics,
University of Toronto, Toronto, M5S 3H8, Canada\\
Department of Astronomy and Astrophysics,
University of Toronto, Toronto, M5S 3H8, Canada}

\begin{document}

\maketitle

\begin{abstract} 

Weak gravitational lensing provides a direct way to study the mass
distribution of clusters of galaxies at large radii. Unfortunately,
large scale structure along the line of sight also contributes to the
lensing signal, and consequently affects the measurements.  We
quantify the effect of distant uncorrelated large scale structure on
the inferred mass profile of clusters as measured from weak
lensing. We consider NFW profiles, and find that large scale structure
is a major source of uncertainty for most practical situations, when a
model, with the mass $M_{200}$ and the concentration parameter $c$ as
free parameters, is fit to the observations. We find that the best
constraints are found for clusters at intermediate redshifts
$(z\approx 0.3)$. For a cluster at $z=0.3$, optimal results are
obtained when the lensing signal is measured out to $10-15$
arcminutes.  Measurements at larger radii do not improve the accuracy
with which the profile can be determined, contrary to what is expected
when the contribution from large scale structure is ignored. The true
uncertainties in $M_{200}$ and the concentration parameter $c$ are
$\sim 2$ times larger than when distant large scale structure is not
included in the error budget.

\end{abstract}


\section{Introduction}

Dark matter plays an important role in the formation of structure in
the universe. However, little is known about the nature of dark
matter, but it has become clear that the simplest paradigm, that of
collisionless cold dark matter (CDM), can explain the observed
structure in the universe quite well. Numerical simulations, which
provide a powerful way to study the formation of structure in the
universe, indicate that CDM gives rise to a particular density profile
(e.g., Dubinski \& Carlberg 1991; Navarro, Frenk, \& White 1995;
Navarro, Frenk, \& White 1997, hereafter NFW; Moore et al. 1999). In
these simulations, the NFW profile appears to be an excellent
description of the radial mass distribution for halos with a wide
range in mass.

The density profile on small scales, however, remains controversial as
other groups find different results (e.g., Ghigna et al. 2000). In
addition, realistic simulations should include the effect of baryons,
which complicate matters even further. On the other hand, there is
good agreement for the density profile on large scales, and a
comparison of the profiles of real objects with the predictions
provides an important test of the assumption that structures form
through dissipationless collapse.

Clusters of galaxies are particularly well suited for such a
comparison, because their formation can be simulated fairly
well. Unfortunately, because clusters are dark matter dominated, it is
difficult to measure the mass distribution at large radii from the
cluster centre. Dynamical estimators can be used, but assumptions have
to be made about the shape of the cluster, its dynamical state, and
the velocity anisotropies (e.g., van der Marel et al. 2000).

A promising approach is that of weak gravitational lensing. The tidal
gravitational field of the cluster distorts the images of distant
background galaxies. This distortion can be measured and can be used
to reconstruct the projected mass density of the cluster (e.g., Kaiser
\& Squires 1993). The advantage of weak lensing over dynamical
methods is that the results do not rely on the dynamical state or
the nature of the deflecting matter. More importantly, it does not
require a tracer of the gravitational potential (which limits many
dynamical methods), and as a result the lensing signal can be measured
out to large radii from the cluster centre.

The weak lensing signal induced by galaxy clusters has been measured
for a large number of systems (e.g., Bonnet, Mellier \& Fort 1994; Fahlman
et al. 1994; Clowe et al. 2000; Hoekstra et al. 2002a; Squires et
al. 1996). These early studies were limited by the field of view of
the CCD cameras used in the observations. With the recent introduction
of wide field imagers it is now possible to measure the lensing signal
out to large radii (Clowe \& Schneider 2001, 2002) or study the mass
distribution in superclusters (e.g., Kaiser et al. 1998; Gray et
al. 2002). Many other weak lensing applications are discussed in
Hoekstra, Yee \& Gladders (2002b).

The lensing signal at large radii is low, and special care has to be taken
to correct for observational distortions, such as PSF anisotropy,
seeing and camera shear. These systematic signals typically have
amplitudes comparable to the lensing signal one is interested
in. Fortunately, detailed studies have shown that the lensing signal
can be recovered with great accuracy (e.g., Hoekstra et al. 1998;
Erben et al. 2001; Bacon et al. 2001).

However, another source of error is usually ignored: the contribution
of large scale structure. The weak lensing signal is sensitive to all
matter along the line of sight. In general, cluster studies assume
that the signal introduced by the cluster is so dominant that the
contributions from other structures can be neglected.

The effect of local large scale structure (such as the filaments
connecting clusters and groups) on cluster mass estimates has been
studied through numerical simulations (Cen 1997; Metzler et al. 1999;
White et al. 2002). These studies show that local structures introduce
both a bias and additional noise in the mass measurement. Estimates of
cluster mass-to-light ratios are affected less by these structures, as
both the mass and the light are expected to probe the same structure.

The study of the contribution from distant large scale structure (not
correlated with the cluster) requires large cosmological simulations,
as one needs to account for all matter along the line of
sight. Hoekstra (2001) demonstrated how the contribution of distant
large scale structure can be computed analytically. Hoekstra (2001)
showed that the large scale structure introduces noise in the cluster
mass measurement, but does not bias the result. The results presented
by White et al. (2002), based on detailed numerical simulations, are
in good agreement with the findings from Hoekstra (2001).

It is important to note that the distant large scale structure does
affect the determination of the cluster mass-to-light ratio.
Furthermore, the distant large scale structure provides a fundamental
limit to the accuracy of weak lensing mass estimates: deeper 
observations reduce the statistical uncertainty (because of the
intrinsic shapes of the sources), but increase the noise from
large scale structure (Hoekstra 2001).

These studies have demonstrated that the large scale structure can be
an important source of noise for the mass determination of individual
objects. The cluster mass is an integral over the cluster profile, and
therefore large scale structure will have an impact on the
determination of cluster mass profiles.

Local large scale structure and substructure in the cluster will
change the profile. This was studied by King, Schneider \& Springel
(2001) using a numerical simulation of a cluster. They found that
substructure increases the error on the parameters by only $\sim 3\%$.
Doubling the amount of substructure resulting in $\sim 10\%$ larger
errors. Based on these results, King et al. (2001) concluded that
weak lensing studies of clusters can be of great use to study
mass profiles.

However, King et al. (2001) did not consider the contribution of
distant large scale structure to the error budget. In this paper, we
use the approach introduced by Hoekstra (2001) to quantify the
contribution of distant large scale structure on measurements of
cluster density profiles. 

The structure of the paper is as follows. In Section~2 we calculate
the contribution of distant large scale structure to the lensing
signal. The lensing properties of the NFW profile are discussed in
Section~3. The effects of distant large scale structure are studied in
Section~4, where we consider clusters with different masses and
redshifts.

\section{Noise from large scale structure}

In the weak lensing regime, the coherent distortion of the images of
the background galaxies provides a direct measure of the lensing shear
$\gamma$ as a function of position. The projected mass distribution of
a cluster can be reconstructed in a parameter-free way from the
observed shear field (e.g., Kaiser \& Squires 1993). The resulting
mass map can be used to derive the cluster mass profile.  However, the
mass reconstruction is a non-local operation, and as a result the
noise in the map varies as a function of position.

Although one can estimate the noise in non-parametric reconstructions
from bootstrap simulations, it is more convenient to use parameterised
models which are fitted to the observations. In the latter case the
error properties are well understood. To study the distribution of
matter in the cluster, 2-dimensional models can be fitted. Another
approach is to consider the azimuthally averaged tangential shear as a
function of radius from the centre.  The latter is particularly useful
to estimate the mass of the cluster (e.g., Hoekstra et al. 1998) or
study the mass profile (e.g., Clowe et al. 2000; Clowe \& Schneider
2001; Clowe \& Schneider 2002; Hoekstra et al. 2002a).

Hoekstra (2001) showed how distant large scale structure affects the
mass estimate when a singular isothermal sphere (SIS) model is fitted
to the tangential shear profile. In this paper we apply the approach
presented in Hoekstra (2001) to estimate how large scale structure
affects the measured tangential shear profile. This allows us to
efficiently simulate observations, and quantify the effect of distant
large scale structure on the derived cluster profile. We focus on the
NFW profile, which is discussed in Section~3, although the procedure
can be readily applied to any model.

The quantity of interest is the tangential shear averaged in
an annulus ranging from $\theta_1$ to $\theta_2$:
$\langle\gamma_T\rangle (\theta_1,\theta_2)$. We will show that the
tangential shear in an annulus is a filtered measurement of the
surface mass density, i.e., it is a particular choice of the aperture
mass statistic $M_{\rm ap}$ (e.g., Schneider et al. 1998). The
aperture mass is defined as

\begin{equation}
M_{\rm ap}(\theta)=\int_{0}^{\theta} d^2\vt U(|\vt|)\kappa(\vt),
\end{equation}

\noindent where $U(\theta)$ is the weight or filter function. Provided
that $U(\theta)$ is compensated, i.e.

\begin{equation}
\int_0^{\theta} d\vt~\vt U(\vt)=0,
\end{equation}

\noindent the aperture statistic can be related to the tangential
shear $\gamma_T$. Unfortunately $\gamma_T$ cannot be observed
directly, but one can measure the distortion
$g_T=\gamma_T/(1-\kappa)$.  In the weak lensing regime, where
$\kappa\ll 1$, one can use $\gamma_T\approx g_T$, which is what we
do throughout this paper. Hence,

\begin{equation}
M_{\rm ap}(\theta)=\int_0^{\theta} d^2\vt Q(|\vt|)\gamma_T(\vt) \label{eqmapq}.
\end{equation}

\noindent $Q(\theta)$ is related to $U(\theta)$ through

\begin{equation}
Q(\theta)=\frac{2}{\theta^2}\int_0^{\theta} d\vt \vt U(\vt) - U(\theta).
\end{equation}

Hence, similar to the mass estimate of the cluster (Hoekstra 2001),
the tangential shear in an annulus can be written as an aperture mass,
with the filter function $Q(\theta)$

\begin{equation}
Q(|\theta|)=\left\{
\begin{array}{ll}
\frac{1}{\pi(\theta_2^2-\theta_1^2)} & (\theta_1\le|\theta|\le \theta_2)\\
 & \\
0 & ({\rm elsewhere})\\
\end{array}
\right. .
\end{equation}

\noindent and the corresponding $U(|\theta|)$ is

\begin{equation}
U(|\theta|)=\left\{
\begin{array}{ll}
\alpha_1 &(\theta<\theta_1)\\
\alpha_2\ln(\theta)+\alpha_3 &(\theta_1\le|\theta|\le \theta_2)\\
0 & (\theta>\theta_2)\\
\end{array}
\right. .
\end{equation}

\noindent The coefficients $\alpha_i$ are defined as 

\begin{equation}
\alpha_1=\frac{2[\ln(\theta_2)-\ln(\theta_1)]}{\pi(\theta_2^2-\theta_1^2)},
\end{equation}

\begin{equation}
\alpha_2= \frac{-2}{\pi(\theta_2^2-\theta_1^2)},
\end{equation}

\noindent and

\begin{equation}
\alpha_3=\frac{2\ln(\theta_2)-1}{\pi(\theta_2^2-\theta_1^2)}.
\end{equation}

Under the assumption that the cluster is the only lensing structure in
the field, the measured $\langle\gamma_T\rangle(\theta_1,\theta_2)$ as
a function of the distance to the cluster centre can be readily used
to study the cluster mass profile: the parameters describing the
cluster can be inferred from a least squares fit of the model to the
data. However, the observed weak lensing signal is sensitive to all
matter along the line of sight, and other structures (the distant
large scale structure) will introduce an additional signal, and 
the observed signal is the sum of the contribution from the cluster
and large scale structure

\begin{equation}
\langle\gamma_T\rangle_{\rm obs}=\langle\gamma_T\rangle_{\rm cluster}+
\langle\gamma_T\rangle_{\rm LSS}.
\end{equation}

\begin{figure}
\begin{center}
\leavevmode
\hbox{%
\epsfxsize=\hsize
\epsffile[35 180 575 700]{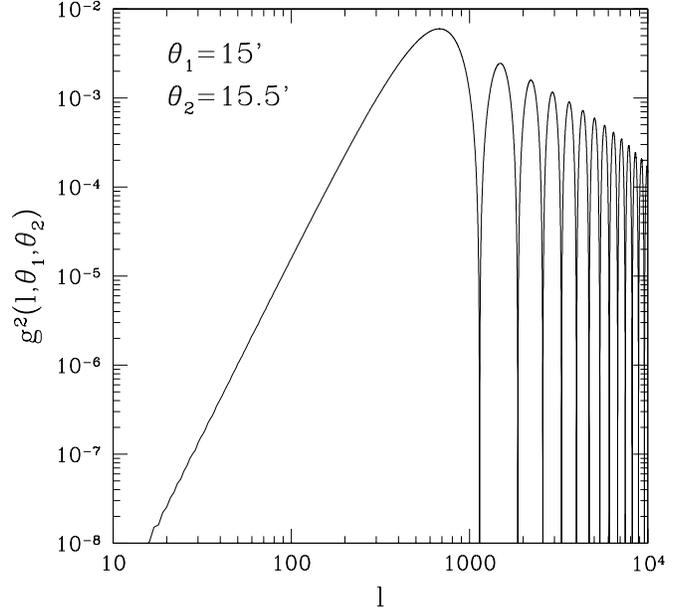}}
\begin{small}      
\caption{The function $g^2(l, \theta_1,\theta_2)$ as a function of
$l$, for $\theta_1=15'$ and $\theta_2=15\farcm 5$. The filter function
declines rather slowly ($\propto 1/l$).\label{filter}}
\end{small}
\end{center}
\end{figure}

The aperture mass measures a density contrast, and therefore the
expectation value for the contribution by uncorrelated large scale
structure vanishes (i.e., $\langle M_{\rm LSS}\rangle=0$). Thus on
average, distant structures do not bias the measurements, but they
introduce and additional statistical uncertainty in the measurement of
size $\sigma^2_{\rm LSS} \equiv \langle\langle\gamma_T\rangle_{\rm
LSS}^2\rangle^{1/2}$.  However, because of non-linear gravitational
evolution, the distribution of $M_{\rm ap}$ will be skewed (e.g.,
Bernardeau et al. 1997; Schneider 1998). In the analysis we will
ignore the skewness, and assume that $M_{\rm ap}$ follows a normal
distribution.

The uncertainty in the measurement of the average tangential shear
$\sigma_{\rm obs}$ in a bin ranging from $\theta_1$ to $\theta_2$ is

\begin{equation}
\sigma^2_{\rm obs}=\sigma^2_{\rm gal}+\sigma^2_{\rm LSS},
\end{equation}

\noindent where $\sigma^2_{\rm gal}$ corresponds to the uncertainty
introduced by the intrinsic shapes of the galaxies and the noise in
the shape measurement. It depends on the number density of sources
$\bar n$ and the scatter in the ellipticities $\sigma_\epsilon$

\begin{equation}
\sigma^2_{\rm gal}=\frac{\sigma_\epsilon^2}{\pi (\theta_2^2-\theta_1^2)\bar n}.
\end{equation}

\noindent For the dispersion in the shapes of the sources we adopt a
value $\sigma_{\rm \epsilon}=0.25$, similar to what is observed in
deep HST images (e.g., Hoekstra et al. 2000). Depending on the image
quality and depth of the observations, the dispersion is likely to be
larger for faint galaxies in ground based data, because of the
correction for the circularization by the PSF. The number density of
galaxies with $20<R<26$ is $\sim 60$ galaxies arcmin$^{-2}$ in space
based observations. In ground based data, however, the effective
number density of galaxies is lower, because many of the faint
galaxies have sizes comparable to the PSF, and are not useable in the
weak lensing analysis (i.e., these galaxies have $\sigma_\epsilon\gg
0.25$). We therefore adopt a typical value of $30$ galaxies
arcmin$^{-2}$ (e.g., Bacon et al. (2002)).

The variance in the aperture mass caused by large scale structure
is given by (Schneider et al. 1998)

\begin{equation}
\sigma^2_{\rm LSS}(\theta_1,\theta_2)= 2\pi\int_0^\infty dl~l
P_\kappa(l) g^2(l,~\theta_1,\theta_2), \label{eqmap}
\end{equation}

\noindent where the effective projected power spectrum $P_\kappa(l)$ is 
given by:

\begin{equation}
P_\kappa(l)=\frac{9 H_0^4 \Omega_m^2}{4 c^4}
\int_0^{w_H} dw \left(\frac{\bar W(w)}{a(w)}\right)^2 
P_\delta\left(\frac{l}{f_K(w)};w\right)
\end{equation}

\noindent Here $w$ is the radial coordinate, $a(w)$ the cosmic scale factor,
and $f_K(w)$ the comoving angular diameter distance. $\bar W(w)$ is the
source averaged ratio of angular diameter distances $D_{ls}/D_{s}$ for
a redshift distribution of sources $p_w(w)$:

\begin{equation}
\bar W(w)=\int_w^{w_H} dw' p_w(w')\frac{f_K(w'-w)}{f_K(w')}.
\end{equation}

\noindent The function $g(l,~\theta_1,\theta_2)$ in
equation~\ref{eqmap} depends on the filter function $U(|\theta|)$ as

\begin{equation}
g(l,~\theta_1,\theta_2)=\int_0^{\theta_2} d\phi\phi U(\phi)J_0(l\phi).
\end{equation}

\begin{figure}
\begin{center}
\leavevmode
\hbox{%
\epsfxsize=\hsize
\epsffile[35 180 575 700]{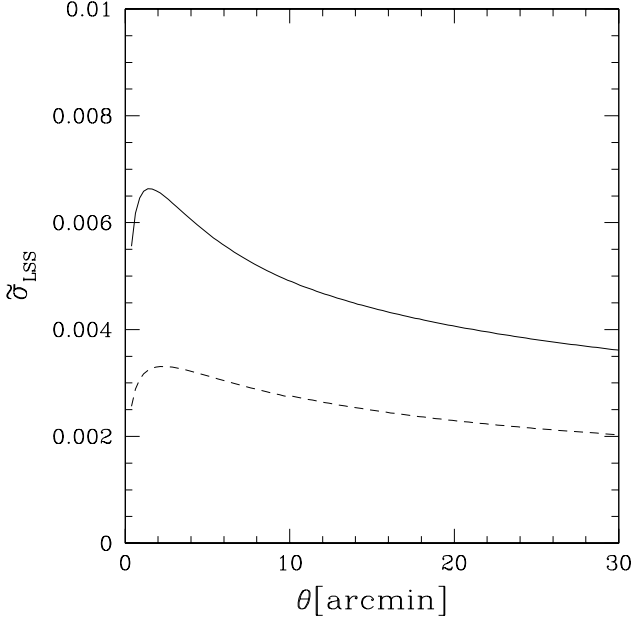}}
\begin{small}      
\caption{The dispersion $\tilde\sigma_{\rm LSS}(\theta)$ in the
averaged tangential shear introduced by distant large scale structure.
The solid lines shows the results for sources with $20<R<26$ and
the dashed line corresponds to sources with $20<R<24$. The effect
is smaller for brighter (lower redshift) sources, but in this
case the cluster lensing signal is lower as well.
\label{gt_lss}}
\end{small}
\end{center}
\end{figure}

As has been demonstrated in Jain \& Seljak (1997) and Schneider et
al. (1998) it is necessary to use the non-linear power spectrum in
equation~14. This power spectrum can be derived from the linearly
evolved cosmological power spectrum following the prescriptions of
Hamilton et al. (1991), and Peacock \& Dodds (1996).

For our choice of filter function $U(|\theta|)$ we obtain

\begin{eqnarray}
& g(l,\theta_1,\theta_2) &= 
\left(\frac{1-2\ln(\theta_1)}{\pi(\theta_2^2-\theta_1^2)}\right)
\frac{\theta_1 J_1(l\theta_1)}{l}
\nonumber\\
& & +\alpha_3\frac{\theta_2 J_1(l\theta_2)}{l}+
\alpha_2\int_{\theta_1}^{\theta_2}d\phi \phi \ln(\phi)J_0(l\phi).
\end{eqnarray}

Figure~\ref{filter} shows $g^2(l,\theta_1,\theta_2)$ as a function of
$l$ for $\theta_1=15'$ and $\theta_2=15\farcm 5$. These results show that
$g^2$ declines rather slowly ($\propto 1/l$) with increasing $l$,
and therefore probes a large range in $l$.

For the analysis in this paper it is convenient to consider the
aperture mass statistic $\tilde\sigma_{\rm LSS}(\theta)$, which corresponds
to the the average tangential shear measured in an annulus
$[\theta-\delta\theta/2,\theta+\delta\theta/2]$. Provided that
$\delta\theta/\theta\ll 1$, one obtains a simple form

\begin{equation}
g(l,\theta,d\theta)\approx g(l\theta)=\frac{J_2(l\theta)}{2\pi},
\end{equation}

\noindent which is independent of $\delta\theta$, and is
an excellent approximation to the exact calculations. In the
remainder of the paper we will adopt a value of $\delta\theta=0\farcm{5}$.

Figure~\ref{gt_lss} shows the dispersion $\tilde\sigma_{\rm
LSS}(\theta)$ in the averaged tangential shear introduced by distant
large scale structure. To derive these results, we have used the
redshift distribution presented in Hoekstra (2001), which are based on
photometric redshift studies of the Hubble deep fields.  The solid
lines shows the results for sources with $20<R<26$ and the dashed line
corresponds to sources with $20<R<24$. 

The number density and mean source redshifts are higher in deeper
observations. Consequently, the lensing signal is higher, and the
statistical noise is lower. However, as Figure~\ref{gt_lss} shows, the
effect of large scale structure is also increased, and the actual
accuracy of the measurements improves only marginally (also see
Hoekstra 2001).

\begin{figure}
\begin{center}
\leavevmode
\hbox{%
\epsfxsize=\hsize
\epsffile[35 180 575 700]{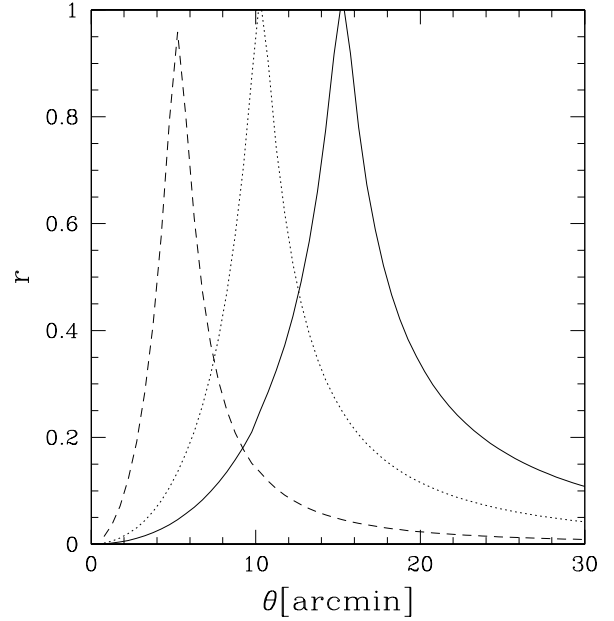}}
\begin{small}      
\caption{The cross-correlation coefficient $r$ for
$\theta=15'$ (solid line), $\theta=10'$ (dotted line)
and $\theta=5'$ (dashed line). The measurements on
at small radii are not strongly correlated, but at
large radii the correlations between bins is rather
strong.
\label{crosscor}}
\end{small}
\end{center}
\end{figure}

To study the effect of distant large scale structure on the
measurement of the cluster mass profile, we create simulated
tangential shear profiles. We include the noise caused by the
intrinsic shapes of the background galaxies, and the noise introduced
by intervening large scale structure.

The statistical error $\sigma_{\rm gal}$ in each bin, caused by the
intrinsic shapes of the sources, is uncorrelated with the other
bins. The noise from large scale structure at various scales is
correlated. To properly account for this, we need to compute not only
$\tilde\sigma_{\rm LSS}$ but also the cross correlation coefficient
between bins. The cross-correlation between bins $\langle
\tilde\sigma_{\rm LSS}(\theta) \tilde\sigma_{\rm
LSS}(\theta-\vt)\rangle$ is calculated using

\begin{equation}
\langle \tilde\sigma_{\rm LSS}(\theta_i)\tilde\sigma_{\rm
LSS}(\theta_j)\rangle= 2\pi\int_0^{\infty}dl~l P_\kappa(l)
g(l,\theta) g(l,\theta-\vt).
\end{equation}

\noindent and the cross-correlation coefficient $r(\theta;\theta-\vt)$ is

\begin{equation}
r(\theta;\theta-\vt)= \frac{\langle \tilde\sigma_{\rm LSS}(\theta)
\tilde\sigma_{\rm LSS}(\theta-\vt)\rangle}
{\sqrt{\langle\tilde\sigma_{\rm LSS}^2(\theta)\rangle
\langle\tilde\sigma_{\rm LSS}^2(\theta-\vt)\rangle}}.
\end{equation}

The cross-correlation coefficients $r$ for three different values of $\theta$
are shown in Figure~\ref{crosscor}. The measurements on small scales
have small range correlations, whereas at large radii the contribution
of large scale structure introduces large scale correlations.

Correlated random noise $c(\theta)$ can be obtained by smoothing of
the white noise $n(\theta)$ with a an appropriate smoothing kernel
$K(\theta-\vt)$

\begin{equation}
c(\theta) = \int  K(\theta-\vt) n(\vt) d\vt, 
\end{equation}

\noindent where $n(\theta)$, $c(\theta)$ are random functions with
correlators

\begin{equation}
\langle n(\theta) n(\theta')\rangle = \delta(\theta-\theta'), 
\end{equation}

\noindent and

\begin{equation}
\langle c(\theta)c(\theta')\rangle= \int K(\theta-\vt)K(\theta'-\vt) d\vt.
\end{equation}

For $n(\theta)$ we take a random number with a Gaussian distribution
with a dispersion of unity. The smoothing kernel is readily obtained
from the cross correlation coefficient. The correlation coefficient
is the convolution of the smoothing kernel with itself. The smoothing
kernel is normalized such that $\langle c^2(\theta)\rangle=
\tilde\sigma^2_{\rm LSS}(\theta)$.

\section{NFW profile}

Numerical simulations have indicated that dark matter halos
originating from dissipationless collapse of density fluctuations can
be described by a universal density profile (e.g., Dubinski \&
Carlberg; Navarro, Frenk, \& White 1995; Navarro, Frenk, \& White
1997). The NFW density profile is given by

\begin{equation}
\rho(r)=\frac{\delta_c\rho_c}{(r/r_s)(1+r/r_s)^2},
\end{equation}

\noindent where $\rho_c$ is the critical density of the universe at the 
redshift of the halo. The overdensity of the halo is parameterized
by $\delta_c$, which is related to the concentration parameter $c$
through

\begin{equation}
\delta_c=\frac{200}{3}\frac{c^3}{\ln(1+c)-c/(1+c)}.
\end{equation}

\noindent The scale radius $r_s$ is the characteric radius of the
halo, which depends on the virial radius $r_{200}$
and the concentration parameter $c$ as $r_s=r_{200}/c$. The virial
radius is defined as the radius where the mass density of the halo
is equal to $200\rho_c$, and the corresponding mass $M_{200}$ inside
this radius is given by

\begin{equation}
M_{200}=\frac{800\pi}{3}\rho_c r_{200}^3.
\end{equation}

Given the cosmology, redshift, and mass of the halo, $r_{200}$ follows
immediately. The simulations have also shown that $c$ and $M_{200}$
are correlated, albeit with some scatter. Hence, given $M_{200}$, the
values of $\delta_c$, and $c$ can be computed using the routine {\tt
CHARDEN} made available by Julio Navarro\footnote{The routine {\tt
CHARDEN} can be obtained from {\tt
http://pinot.phys.uvic.ca/${\tilde{\ }\!}$jfn/charden}}. In this case
the halo profile is described by a single parameter, e.g., the
mass. However, one can also consider $c$ as a free parameter in the
model. Consequently comparison of the inferred value with the
predictions of the numerical simulations provides a direct
observational test of our current understanding of structure
formation.

For the NFW profile, the tangential shear $\gamma_T$ as a function of
radius $\theta$ is given by (Bartelmann 1996; Wright \& Brainerd 2000)

\begin{equation}
\gamma_T(\theta)=\left\{
\begin{array}{ll}
\frac{r_s\delta_c\rho_c}{\Sigma_c} g_<(x) &, x<1\\
 & \\
\frac{r_s\delta_c\rho_c}{\Sigma_c}\left[\frac{10}{3}+4\ln\left(\frac{1}{2}\right)\right] &, x=1\\
 & \\
\frac{r_s\delta_c\rho_c}{\Sigma_c} g_>(x) &, x>1\\
\end{array}
\right. ,
\end{equation}

\noindent where $x=\theta/r_s$. The critical surface density
$\Sigma_c$ is given by

\begin{equation}
\Sigma_c=\frac{c^2}{4\pi G}\frac{D_s}{D_l D_{ls}},
\end{equation}

\noindent where $D_l$, $D_s$, and $D_{ls}$ are, respectively, the
angular diameter distances between the observer and the lens, the
observer and the source, and the lens and the source.  The functions
$g_<(x)$ and $g_>(x)$ are defined as

\begin{eqnarray}
 & g_{<}(x)= & \frac{8{\rm arctanh} \sqrt{\frac{1-x}{1+x}}}{x^{2}\sqrt{1-x^{2}}} 
+ \frac{4}{x^{2}}\ln\left(\frac{x}{2}\right) - \frac{2}{\left(x^{2}-1\right)} \nonumber \\
 &  & +\frac{4{\rm arctanh}\sqrt{\frac{1-x}{1+x}}}{\left(x^{2}-1\right)\left(1-x^{2}\right)^{1/2}},
\end{eqnarray}

\noindent and

\begin{eqnarray}
 & g_{>}(x)= & \frac{8\arctan\sqrt{\frac{x-1}{1+x}}}{x^{2}\sqrt{x^{2}-1}}
+\frac{4}{x^{2}}\ln\left(\frac{x}{2}\right) - \frac{2}{\left(x^{2}-1\right)}\nonumber\\
 &  & +\frac{4\arctan\sqrt{\frac{x-1}{1+x}}}{\left(x^{2}-1\right)^{3/2}}.
\end{eqnarray}

\begin{figure}
\begin{center}
\leavevmode
\hbox{%
\epsfxsize=\hsize
\epsffile[35 180 575 700]{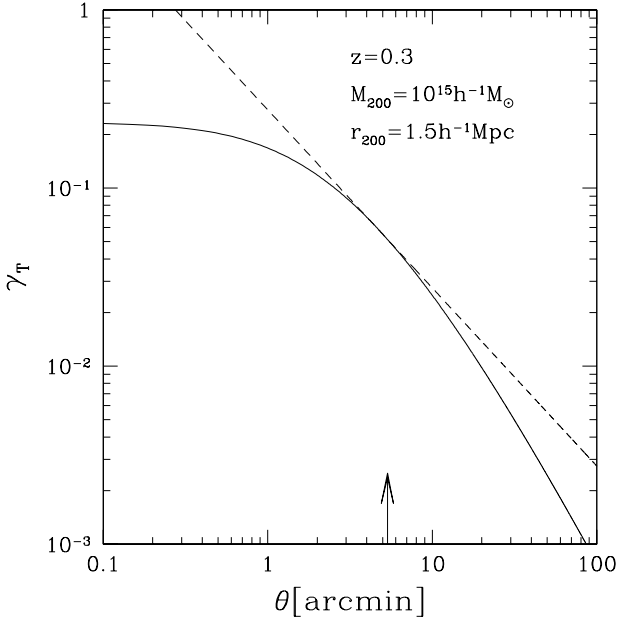}}
\begin{small}      
\caption{Comparison of the tangential shear for a cluster with a NFW
profile (solid line), and a cluster with a SIS profile (dashed line).
The NFW cluster has a mass $M_{200}=10^{15} h^{-1} M_\odot$, and is at
a redshift $z=0.3$. The amplitude of the signal corresponds to that of
source galaxies with apparent magnitudes $20<R<26$. The amplitude of
the SIS profile was matched to the NFW profile at intermediate
scales. The arrow indicates a radius of $1 h^{-1}$ Mpc.
\label{nfw}}
\end{small}
\end{center}
\end{figure}

Figure~\ref{nfw} shows the tangential shear for a cluster with an NFW
profile. The cluster has a mass $M_{200}=10^{15} h^{-1} M_\odot$, and
is at a redshift $z=0.3$. The amplitude of the lensing signal
corresponds to that for source galaxies with apparent magnitudes
$20<R<26$. The dashed line in Figure~\ref{nfw} corresponds to a
Singular Isothermal Sphere (SIS) model, which was matched to the NFW
model at intermediate radii.

In principle one can distinguish between the SIS model and the NFW
profile by examining the lensing signal on either small or large
scales. In practice, however, substructure in the cluster core, and
contamination of the sample of background galaxies by cluster members,
complicates the small scale comparison. Furthermore, the profile on
small scales is still debated (e.g., Ghigna et al. 2000). The lensing
signal at large radii can be measured accurately, provided that the
shapes have been accurately corrected for observational distortions.

\begin{figure}
\begin{center}
\leavevmode
\hbox{%
\epsfxsize=\hsize
\epsffile[35 180 575 700]{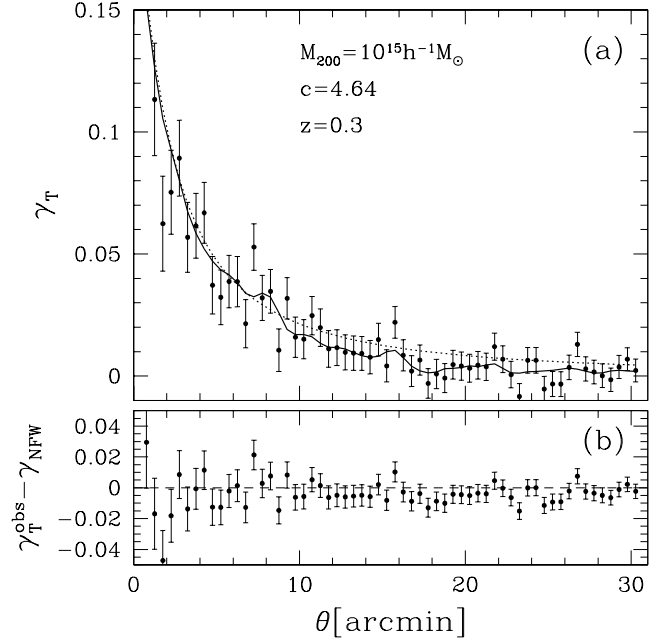}}
\begin{small} 
\caption{(a) Realisation of the observed tangential shear as a
function of radius from the cluster centre for a cluster at a redshift
$z=0.3$ and mass $M_{200}=10^{15}h^{-1}M_\odot$. The points include
the noise caused by the intrinsic shapes of the background galaxies
and the contribution from large scale structure. The error bars on the
points indicate the (uncorrelated) statistical error in the
measurements due to the shapes of the sources (adopting
$\sigma_\epsilon=0.25$ and $\bar n=30$ arcmin$^{-2}$. The model NFW
profile is indicated by the dashed line. The cluster profile with the
contribution from large scale structure added to the signal is
indicated by the solid line. (b) The difference between the observed
signal and the input NFW model. The large scale structure introduces
significant deviations from the model.\label{simnfw}}
\end{small}
\end{center}
\end{figure}

\section{Constraining mass profiles}

To study cluster mass profiles, it is common to compare a parametrized
mass model to the data. Ideally one would like to minimize the number
of parameters describing the model.  In this section we consider the
NFW profile (Eqn.~24), with $M_{200}$ and $c$ as the only two free
parameters. In addition to these two parameters, one could introduce
the slopes of the profile on small and large scales.

To examine how well we can constrain the values of $M_{200}$ and $c$,
we create simulated tangential shear profiles as described in
Section~2. The profiles include the statistical noise due to the
intrinsic shapes of galaxies and the contribution from distant large
scale structure. A typical realisation is presented in
Figure~\ref{simnfw}a.  The points indicate the simulated measurements,
and the error bars reflect only the statistical error (the intrinsic
shapes of the sources).  The input NFW signal is indicated by the
dashed line. The actual lensing signal, which is modified because of
the large scale structure, is given by the solid line

Figure~\ref{simnfw}b shows the difference between the ``observed''
signal and the input NFW profile. The large scale structure introduces
significant deviations, which, in this particular case, systematically
lower the signal on scales larger than 10 arcminutes. The input
profile has $M_{200}=10^{15}h^{-1} M_\odot$ and $c=4.6$. The best fit
NFW profile to the profile with the contribution from large scale
structure only (i.e., no statistical noise from the shapes of the
sources) yields $M_{200}=8.7\times 10^{14} M_\odot$ and $c=5.1$ when
fitted out to 15 arcminutes. When the profile is fitted out to 30
arcminutes, we obtain $M_{200}=7.4\times 10^{14} M_\odot$ and
$c=5.5$. Hence, the large scale structure alters the inferred profile
of an individual cluster, and the inferred values depend on the range
in radius included in the fit. Note, however, that the ensemble
average value is not changed when a large sample of clusters is used.

\begin{figure*}
\begin{center}
\leavevmode
\hbox{%
\epsfxsize=\hsize
\epsffile[35 320 575 670]{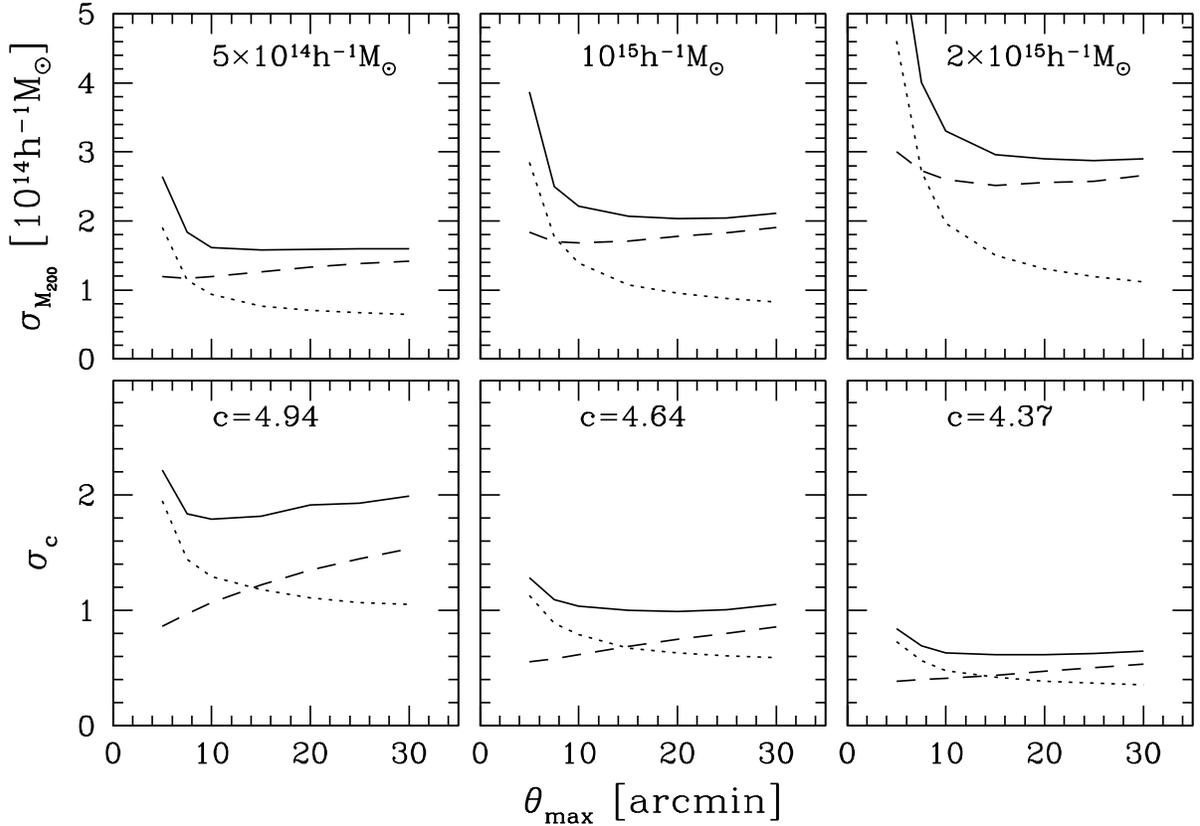}}
\begin{small}      
\caption{Measurement error in the determination of $M_{200}$ (upper
panels) and $c$ (lower panels) as a function of the maximum fitted
radius $\theta_{\rm max}$ for three clusters at a redshift of
$z=0.3$. The results are obtained by fitting an NFW profile to the
simulated tangential shear profile, with $M_{200}$ and $c$ as free
parameters. The uncertainties are determined from 1000 realisations.
The solid lines indicate the total uncertainty, whereas the dashed
line denotes the contribution by large scale structure, and the dotted
line the contribution from the intrinsic ellipticities of the
background galaxies. The results demonstrate that the real uncertainty
is a factor $\sim 2$ larger than expected from the noise caused by the
intrinsic shapes of the sources alone. Also the accuracy in the
parameter estimation does not improve by fitting the model to
radii larger than $\sim 15$ arcminutes.
\label{noise_nfw}}
\end{small}
\end{center}
\end{figure*}

We now quantify the contribution of distant large scale structure to
the total error budget. We start by considering clusters at a redshift
$z=0.3$, and masses $M_{200}=0.5,1~{\rm and}~2\times 10^{15} h^{-1}
M_\odot$.  We assume that, because of substructure in the cluster core
and contamination by cluster members, the NFW profile is fitted at radii
larger than 30 arcseconds. The inclusion of measurements at smaller
radii does not change the results significantly, because of the large
statistical error on small scales.

Figure~\ref{noise_nfw} shows how the $1\sigma$ uncertainty in the
determination of $M_{200}$ (upper panels) and $c$ (lower panels)
varies with increasing $\theta_{\rm max}$, the outermost point
included in the fit. The uncertainties are determined from 1000
realisations. The dotted lines corresponds to the contribution from the
intrinsic ellipticities of the background galaxies. The statistical
uncertainty decreases when the lensing signal is measured out to
larger distances from the cluster centre.  The dashed lines denotes the
contribution from the large scale structure. The combined uncertainty
is indicated by the solid lines.

The results presented in Figure~\ref{noise_nfw} demonstrate that
contribution of distant large scale strucure to the total error budget
cannot be neglected. This is true, even for clusters as massive as
$M_{200}=2\times 10^{15}h^{-1} M_\odot$. The accuracy in the
measurement of $c$ improves with increasing mass of the cluster,
whereas the uncertainty in $M_{200}$ increases (although the relative
accuracy does improve). 

\begin{figure*}
\begin{center}
\leavevmode
\hbox{%
\epsfxsize=8cm
\epsffile{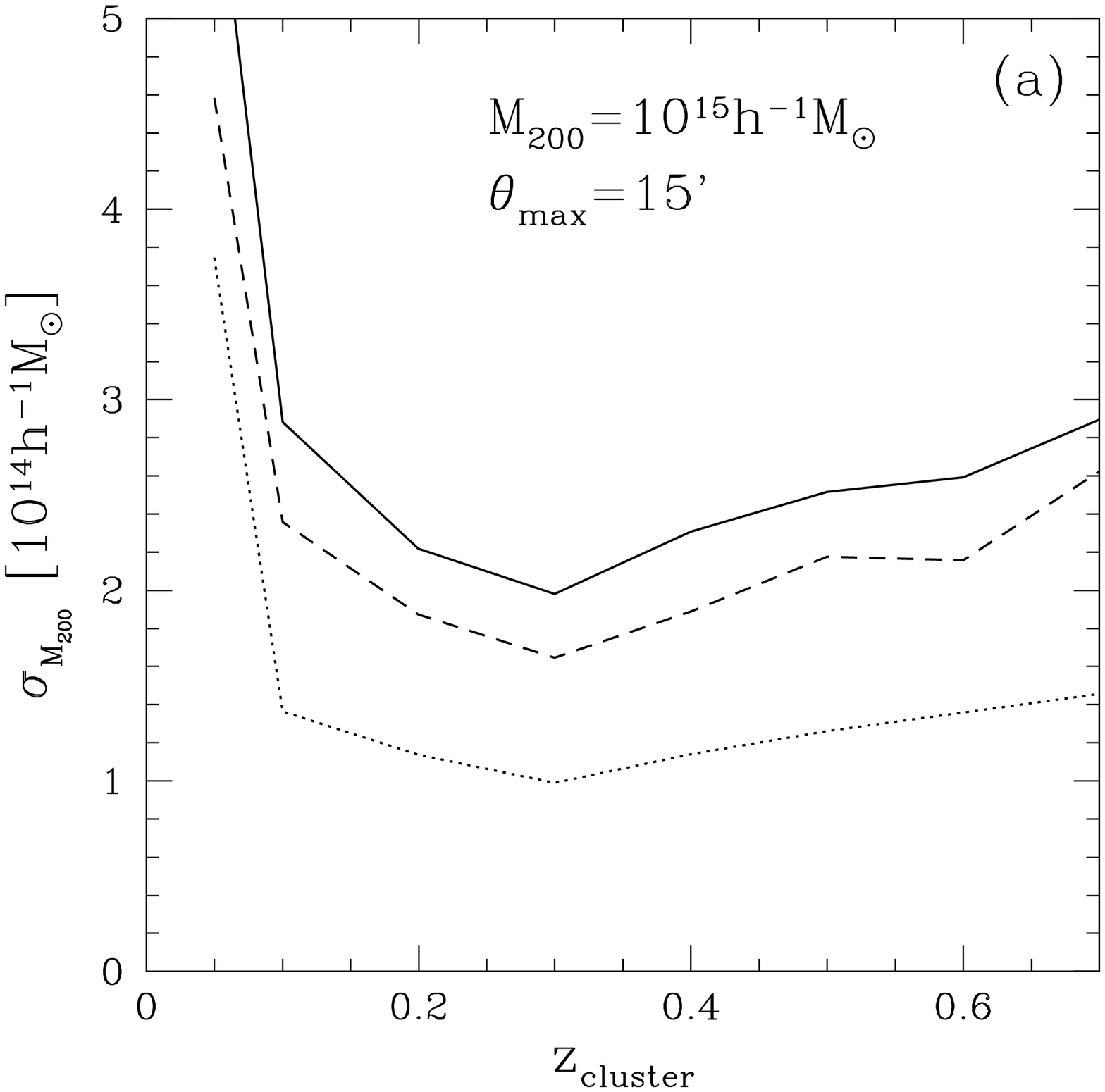}
\epsfxsize=8cm
\epsffile{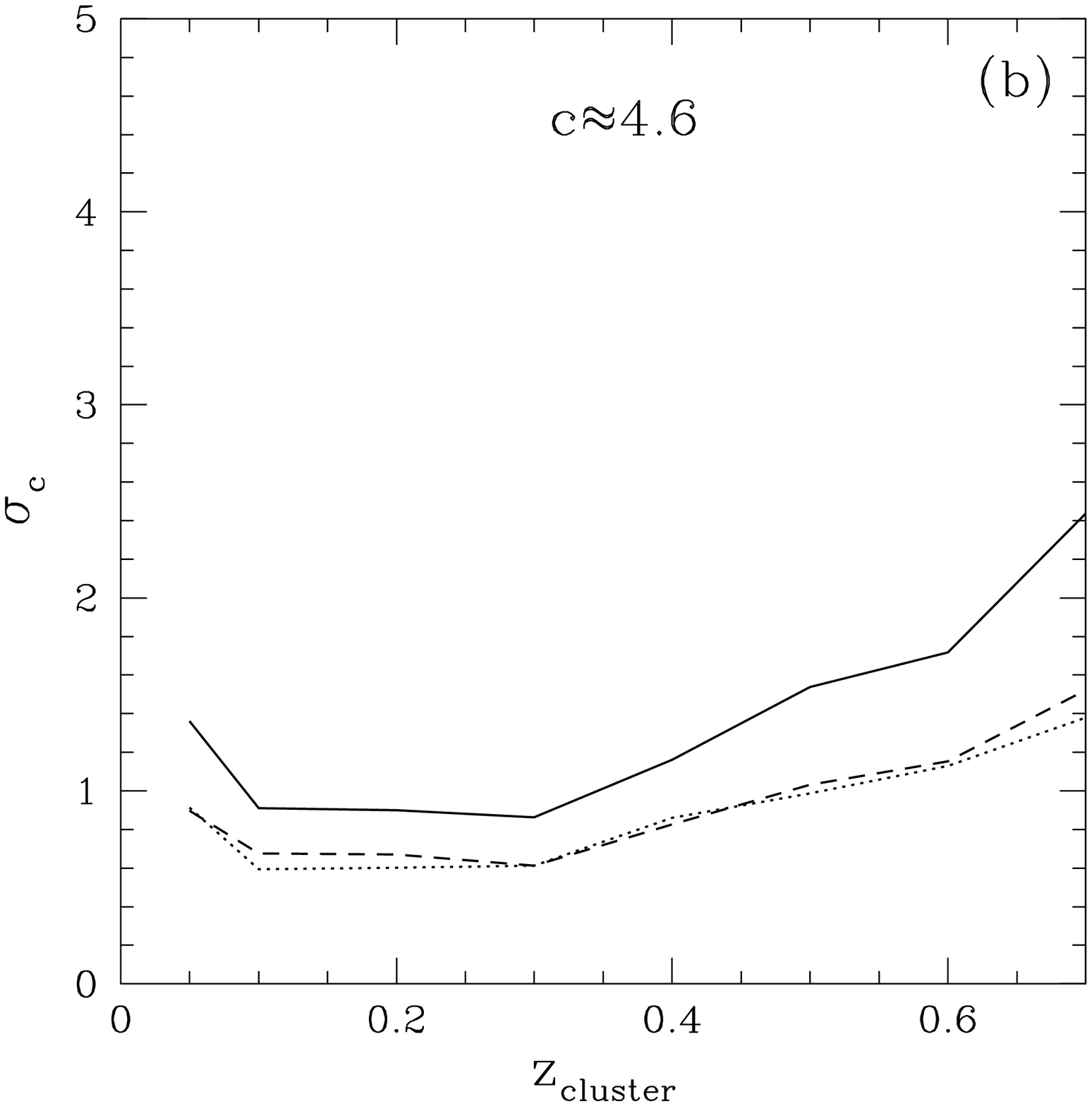}}
\begin{small}      
\caption{Measurement error in the determination of $M_{200}$ (panel a)
and $c$ (panel b) as a function of redshift for a cluster with
$M_{200}=10^{15}h^{-1}$M$_\odot$, if an NFW profile is fit to the data
with $M_{200}$ and $c$ as free parameters. The solid lines indicate
the total uncertainty, whereas the dashed line denotes the
contribution by large scale structure, and the dotted line the
contribution from the intrinsic ellipticities of the background
galaxies. The results are for a maximum fitted radius $\theta_{\rm
max}=15$ arcminutes. The optimal redshift for the determination of
$M_{200}$ and $c$ is $z\sim 0.3$. The uncertainty in $M_{200}$ 
increases slowly with redshift, whereas the uncertainty in $c$
for high redshift clusters is large.
\label{noise_z}}
\end{small}
\end{center}
\end{figure*}

The real uncertainties in $M_{200}$ and $c$ are typically a factor
$\sim 2$ larger than expected from the statistical noise caused by the
intrinsic shapes of the sources alone. Furthermore, the accuracy of
the parameter estimation does not improve by fitting the model to
radii larger than $\sim 15$ arcminutes. 

It is also interesting to examine how well we constrain $M_{200}$ and
$c$ for clusters at various redshifts. For instance, Hoekstra
(2001) showed that large scale structure seriously limits the
accuracy of mass determinations for low redshift clusters. At
higher redshifts, the accuracy is limited by the intrinsic shapes
of the sources. 

We consider a cluster with $M_{200}=10^{15}h^{-1}M_\odot$ at redshifts
ranging from 0.05 to 0.7. Figure~\ref{noise_nfw} shows that the most
accurate results are obtained for $\theta_{\rm max}=15'$, and we use
this value to fit the profiles. The $1\sigma$ uncertainties in the
measurements of $M_{200}$ and $c$ as a function of redshift are
presented in Figure~\ref{noise_z}.

The results show that the optimal redshift for the
determination of $M_{200}$ is $z\sim 0.3$, although for the adopted
limiting magnitude of $R=26$, the uncertainty does not increase
dramatically for higher redshift clusters. This is because for a fixed
angular scale, one probes larger physical scales at high redshift,
which explains why the contribution from the intrinsic shapes 
is almost constant, and the noise from large scale structure
increases. The noise, however, increases rapidly for low redshift
clusters. 

The situation is different for the measurement of the concentration
parameter $c$, which is presented in Figure~\ref{noise_z}b. The
error increases significantly for high redshift clusters, and
the most accurate measurements are obtained for clusters with redshifts
$z=0.1-0.3$. 

Therefore studies of intermediate redshift clusters are best suited to
test the predictions of dissipationless collapse. In addition these
systems can be studied efficiently using X-ray telescopes and new
multi-object spectrographs (such as IMACS). Comparison of the results
obtained through these various techniques can provide also new insights
in the dynamics of structure formations

\section{Conclusions}

The comparison of measured mass profiles of cluster of galaxies
with the outcome of numerical simulations provides an important
test of our current understanding of structure formation. Although
mass profiles can be inferred through various techniques, weak
lensing is of particular interest, because it probes the 
(dark) matter distribution directly. It does not rely on assumptions
about the dynamical state of the cluster or the nature of dark matter.
In addition it does not require a visible tracer, which is
important to study the mass profile at large distances from the
cluster centre.

Unfortunately, a simple interpretation of the lensing signal is
complicated because of the contribution of large scale structure along
the line of sight. To date, weak lensing studies of clusters of
galaxies have ignored the effect of distant large scale structure.
The effect on mass estimates was studied previously by Hoekstra
(2001). In this paper we followed the approach of Hoekstra (2001) to
quantify the effect of distant large scale structure on the mass
profile inferred from weak lensing.

We considered NFW profiles, which provide a good description of the
outcome of simulations with collisionless dark matter. We have created
shear profiles, which include the noise from the intrinsic shapes of
the galaxies and the contribution from large scale structure. These
profiles were fit with NFW profiles, with the mass $M_{200}$ and the
concentration parameter $c$ as free parameters.

We found that $M_{200}$ and $c$ are constrained best for clusters at
intermediate redshifts ($z\approx 0.3$). For a cluster at $z=0.3$,
optimal results are obtained when the lensing signal is measured out
to $10-15$ arcminutes. Measurements at larger radii do not improve
the accuracy. This is different from the situation where the
contribution from large scale structure is ignored: the measurement
error decreases when the signal is measured out to larger radii.

King et al. (2001) studied the effect of substructure on the
determination of cluster mass profiles, and found that the errors in
the fitted parameters increase by a few percent. We found, however,
that distant large scale structure is a major source of uncertainty in
the determination of cluster mass profiles from weak lensing.  The
true uncertainties in $M_{200}$ and the concentration parameter $c$
are $\sim 2$ times larger than when large scale structure is not
included in the error budget. It is also important to note that
deeper observations do not significantly improve the accuracy of
the measurements, because the smaller statistical error is
counteracted by an increase in noise due to large scale structure.

\section*{Acknowledgements}

I thank Kris Blindert, Ludo van Waerbeke and Howard Yee for a careful
reading of the manuscript.

\end{document}